\documentclass[sigconf,review=false]{acmart}
\usepackage{hyperref}
\usepackage{float}
\usepackage{geometry}
\usepackage{multirow}
\usepackage{hyperref}
\usepackage{rotating}
\usepackage{xcolor}
\usepackage{pdflscape}
\usepackage{hyperref}
\usepackage{graphicx}
\usepackage[]{mdframed}
\usepackage{lipsum}
\usepackage{caption}
\usepackage{subcaption}
\usepackage{float}
\usepackage{wrapfig}
\usepackage{booktabs}
\usepackage{tabularx}
\usepackage[utf8]{inputenc}
\usepackage[resetlabels,labeled]{multibib}
\newcites{S}%
{Primary Studies} 
\AtBeginDocument{%
  \providecommand\BibTeX{{%
    \normalfont B\kern-0.5em{\scshape i\kern-0.25em b}\kern-0.8em\TeX}}}

\setcopyright{rightsretained}
\copyrightyear{2024}
\acmYear{2024}
\acmDOI{10.1145/3641822.3641863} 
\acmConference[CHASE '24]{2024 IEEE/ACM 17th International Conference on Cooperative and Human Aspects of Software Engineering }{April 14--15, 2024}{Lisbon, Portugal}
\acmBooktitle{2024 IEEE/ACM 17th International Conference on Cooperative and Human Aspects of Software Engineering (CHASE '24), April 14--15, 2024, Lisbon, Portugal}
\acmISBN{979-8-4007-0533-5/24/04}

\begin{document}

\title{Hybrid Work meets Agile Software Development: A Systematic Mapping Study}

\author{Dron Khanna}
\orcid{0000-0003-4760-5560}

\affiliation{%
  \institution{Free University of Bozen-Bolzano}
  \streetaddress{Piazza Università, 1,}
  \city{Bolzano}
  \country{Italy}
  \postcode{39100}
}
\email{dron.khanna@unibz.it}

\author{Emily Laue Christensen}
\orcid{0000-0002-5678-0179}
\affiliation{%
  \institution{Lappeenranta-Lahti University of Technology}
  \streetaddress{Mukkulankatu 19, 15210}
  \city{Lahti}
  \country{Finland}}
\email{emily.christensen@lut.fi}

\author{Saagarika Gosu}
\orcid{0000-0003-4647-2508}
\affiliation{%
  \institution{Free University of Bozen-Bolzano}
  \streetaddress{Piazza Università, 1,}
  \city{Bolzano}
  \country{Italy}}
\email{saagarika.gosu@student.unibz.it}

\author{Xiaofeng Wang}
\orcid{0000-0001-8424-419X}
\affiliation{%
  \institution{Free University of Bozen-Bolzano}
  \streetaddress{Piazza Università, 1,}
  \city{Bolzano}
  \country{Italy}}
\email{xiaofeng.wang@unibz.it}

\author{Maria Paasivaara}
\orcid{0000-0001-7451-7772}
\affiliation{%
  \institution{Lappeenranta-Lahti University of Technology}
  \streetaddress{Mukkulankatu 19, 15210}
  \city{Lahti}
  \country{Finland}}
\email{maria.paasivaara@lut.fi}

\renewcommand{\shortauthors}{Dron Khanna, et al.}

\begin{abstract}
 Hybrid work, a fusion of different work environments that allow employees to work in and outside their offices, represents a new frontier for agile researchers to explore. However, due to the nascent nature of the research phenomena, we are yet to achieve a good understanding of the research terrain formulated when hybrid work meets agile software development. This systematic mapping study, we aimed to provide a good understanding of this emerging research area. The systematic process we followed led to a collection of 12 primary studies, which is less than what we expected. All the papers are empirical studies, with most of them employing case studies as the research methodology. The people-centric nature of agile methods is yet to be adequately reflected in the studies in this area. Similarly, there is a lack of a richer understanding of hybrid work in terms of flexible work arrangements. Our mapping study identified various research opportunities that can be explored in future research.
\end{abstract}


\maketitle

\section{Introduction}
Hybrid work is a setting where employees have the convenience to work both in the office space and in remote environments (i.e., home, outside office space) \cite{conboy2023future}. According to a recent article by Forbes\footnote{https://www.forbes.com/sites/chriswestfall/2023/07/05/over-95-of-workers-say-that-hybrid-work-is-best-for-mental-health/?sh=280bcba11b4d} more than 95\% of workers believe that hybrid work arrangements are better for their mental health. The same survey outlines that women prefer hybrid work to men. A psychologically and mentally safe working environment should be mandated in every organization \cite{edmondson2021psychological}. If employees achieve better results with online or remote work, the practice should be encouraged \cite{khanna2022your}. Hybrid work is beneficial as it provides flexibility, saves the travel time to commute to the office, promotes the use of collaboration tools with effective communication practices, and enhances work-life balance \cite{nguyen2023work}.

Agile software development is used by many teams as it is a customer-focused approach \cite{flora2014systematic}, where software is delivered fast to market in short delivery sprints \cite{khanna2022know}. Agile software development also provides transparency to projects, enhancing collaboration and communication through cross-functional teams \cite{dikert2016challenges}. The work environment of teams in turn affects their development process, resulting in speeding up or slowing down the development \cite{khanna2018mvps}.
To obtain an overview and understand the research prospects of hybrid work in agile software development organisations, we conducted a Systematic Mapping Study (SMS). An SMS helps to pinpoint areas and gaps that future research could address \cite{petersen2015guidelines}. In addition, it helps to summarise and formulate the current research scope \cite{petersen2008systematic}. The systematic study of the convergence of hybrid work and agile software development provides insights into how organizations manage workflows, teams, online tools, processes, and established work policies. 

As the nature of different types of work is evolving more after the Covid-19 pandemic \cite{nguyen2023work} and several flexible workflows have been flourishing \cite{smite2023work}, a systematic study could offer research insights by showing several future trends, emerging technologies, and transformation, and providing an understanding of hybrid work and agile software development. We address the following research questions in this study.\\\\
\fbox{\begin{minipage}{\dimexpr\columnwidth-2\fboxsep-2\fboxrule\relax}
\textbf{RQ1.) What are the publication trends and characteristics of existing research on hybrid work in agile software} \\\textbf{development?}\\
RQ1.1.) What are the publication years and types of research articles on hybrid work in agile software development?\\
RQ1.2.) Which research methods have been employed in the \\ published studies for hybrid work in agile software development?\\
RQ1.3.) In which countries and organizations has the research been carried out for hybrid work in agile software development?\\
\textbf{RQ2.) Which research questions have been investigated in hybrid work in agile software development?\\
RQ3.) In which kind of hybrid settings is agile software \\ development carried out?}
\end{minipage}}\\\\
 
 To answer the above research questions we performed a systematic search on three well-known academic databases (\href{https://dl.acm.org}{ACM Digital Library}, \href{https://ieeexplore.ieee.org/Xplore/home.jsp}{IEEE Xplore}, and \href{https://www.scopus.com/search/form.uri?display=basic#basic}{Scopus}). We used "Hybrid" AND "Agile" AND "Software development" as three main keywords and inserted OR logic between their synonyms. Following the guidelines of \cite{petersen2008systematic,petersen2015guidelines}, we obtained 3191 studies, which were categorised into ‘accept or yes’, ‘maybe’, ‘reject or no’ and
‘duplicate’. Seven relevant studies were selected after applying the inclusion and exclusion criteria. In addition, we used forward and backwards snowballing on the seven papers, which resulted in an additional 485 papers. Again, after applying our inclusion and exclusion criteria, we found five more relevant articles from forward snowballing and none from the backward search. In total, we discovered 12 relevant studies. All the papers are empirical studies, with most of them employing case studies as the research methodology. Our mapping study identified various research opportunities that can be explored in future research.

In Section 2 we provide a brief overview of the background, including the conceptual framework and model we have applied in this study. Section 3 describes the research steps followed. We present the results in Section 4. Finally, we discuss our findings in Section 5, and provide concluding remarks in Section 6.

\section{Background}
After Covid-19, the working model for many has changed to work-from-home (WFH), work-from-anywhere (WFX) \cite{gusain2020work},
remote work \cite{topp20224}, or hybrid work \cite{conboy2023future}. The term hybrid has recently gained popularity as a catch-all for a number of phrases used in the workplace, and the term is often used and interpreted in different ways. In this study, we use the definition proposed by Conboy et al. \cite{conboy2023future}:\\\\
\fbox{\begin{minipage}{\dimexpr\linewidth-2\fboxsep-2\fboxrule\relax} 
\textit{``Hybrid software development is where some team members work mostly or completely from home, others mostly or completely from the traditional office, and others in some combination of the two—not quite distributed and not quite co-located but, rather, individuals working from anywhere and touching base with the office}\\ \textit{intermittently''.}\end{minipage}}\\\\
This definition excludes work settings where all individuals work either exclusively from the office, or only remotely. 

As far as we are aware of, there is no existing work that offers a good overview of the research area at the intersection of hybrid work and agile software development. This is the primary motivation for us to conduct this mapping study. 

\begin{figure}[H]
    \centering
    \includegraphics[width=\columnwidth]{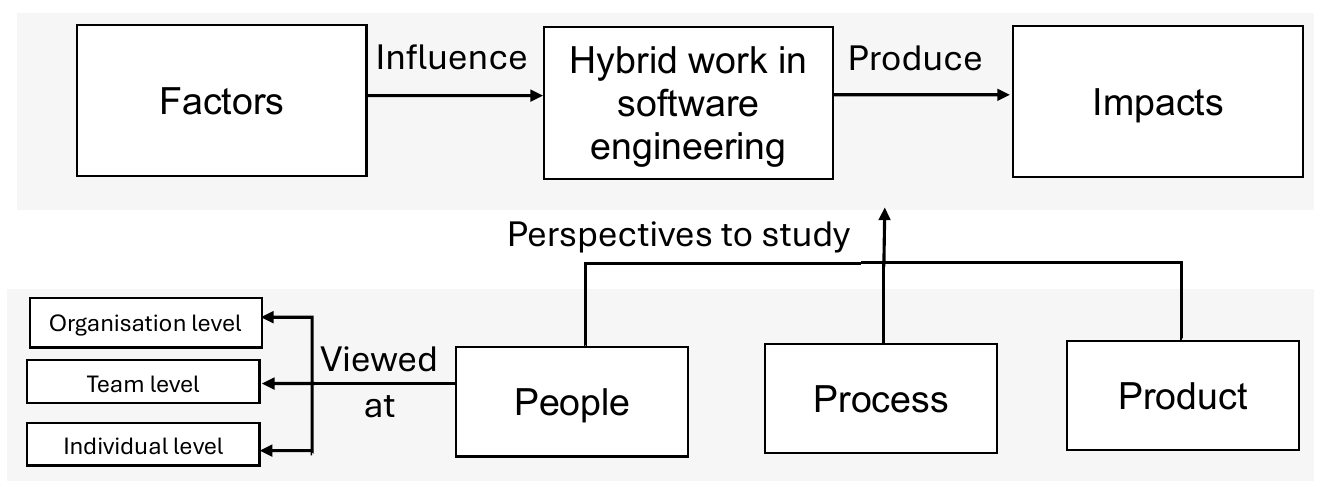}
    \caption{A conceptual framework for organizing research questions on hybrid work in SE \cite{paas}}
    \label{fig:3p}
\end{figure}

\begin{figure*}[t]
    \centering
    \includegraphics[width=1.0\linewidth]{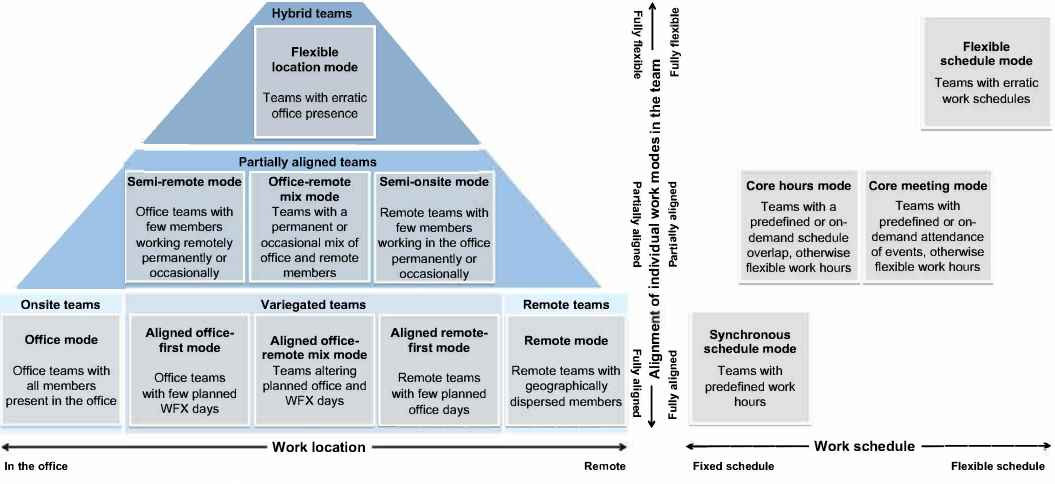}
    \caption{Team typology and the spectrum of work arrangements \cite{smite2022future}}
    \label{fig:wa}
\end{figure*}

To analyse the research questions investigated in the papers selected for this study, and to answer RQ2, we chose the framework suggested by Paasivaara and Wang \cite{paas} and illustrated in Figure \ref{fig:3p}. This framework has been used to organise research topics on hybrid work in Software Engineering (SE) \cite{paas}. 
To further categorize the hybrid settings investigated in these studies, and to answer RQ3, we employed a model suggested by Smite et al. \cite{smite2022future}, that is shown in Figure \ref{fig:wa}. This model suggests various types of hybrid settings in terms of team typology and work arrangements. 

\subsection{Framework for organising research questions}
According to Paasivaara and Wang \cite{paas}, research on hybrid work in SE can be organised into three different categories: 
\begin{itemize}
    \item Those that investigate the \textit{factors} that influence various policies and implementations of hybrid work: 
    \item Those that examine hybrid work arrangements and policies (\textit{hybrid work in SE}): 
    \item Those that analyse the \textit{impacts} of hybrid work.
\end{itemize}

These three categories of research topics can be studied from different perspectives. The three typical perspectives are \textit{People}, \textit{Process}, and \textit{Product}, the so-called ``3 P's'' in software management \cite{Reifer2006}. The \textit{people} perspective can be further divided into \textit{individual, team,} and \textit{organisational} levels.

To make the framework more concrete, a study that investigates \textit{factors} influencing the adoption of hybrid work can do so from the \textit{people} perspective at the \textit{individual level}, studying what personal factors might influence the choice of work mode, e.g., personality or cultural background differentiated by a power distance. 
A study that examines hybrid work arrangements in software companies from the 
\textit{process} perspective could be, for example, understanding how agile methods can be adjusted for hybrid work. Studies on which tools can be used and how they can best support hybrid work in SE also fit to the same category.

To study the impacts of hybrid work on an organisation from the \textit{product} perspective, an example is given in \cite{paas}: would Conway's Law \cite{kwan2011conway} still be applicable in hybrid work? If yes, it would be relevant to investigate how it would be manifested in such settings.

\subsection{Model for analysis of hybrid work settings}
To achieve clarity in the understanding of arrangements of software teams working in hybrid settings, Smite et al. \cite{smite2022future} propose utilizing a team typology and multidimensional model of work arrangements. The two core dimensions of the model are \textit{work location} and \textit{work schedule} (Figure \ref{fig:wa} -- horizontal axis), and the various arrangements within these two dimensions are categorized according to their degree of alignment (Figure \ref{fig:wa} -- vertical axis). The typology provides a vocabulary for the work arrangements of teams specifically. It is, therefore, only applied in this study to categorize the hybrid settings in studies where team work arrangements have been explored.

Similar to the definition of hybrid provided in \cite{conboy2023future}, within the work location dimension, the model distinguishes between arrangements where all individuals are working either fully onsite or fully remote. The proposed definition of \textbf{hybrid} teams is, however, limited to work arrangements where individuals in the team have an erratic office presence with no alignment (\textit{flexible location mode}). 

The remaining types of teams are defined as either: 1) \textbf{partially aligned}, which surfaces when members do not always align, or when everybody's arrangements are not aligned, or 2) \textbf{variegated}, which are distinguished by predefined but altering work locations. Within each of these two degrees of alignment, three different archetypes of work modes are defined: 1) \textit{semi-remote mode}, \textit{office-remote mix mode}, and \textit{semi-onsite mode}, and 2) \textit{aligned office-first mode}, \textit{aligned office-remote mix mode}, and \textit{aligned remote-first mode}.  

Regarding the work schedule dimension, the model distinguishes between four different modes: 1) \textit{synchronous mode} with fully aligned work hours, 2) \textit{flexible mode} with erratic work schedules, 3) \textit{core hours mode} with specific time overlaps, but otherwise flexible schedules, and 4) \textit{core meeting mode} with alignment around scheduled meetings or events, but otherwise flexible schedules.

\section{Research steps}
\begin{figure*}[t]
    \centering
    \includegraphics[width=1.0\linewidth]{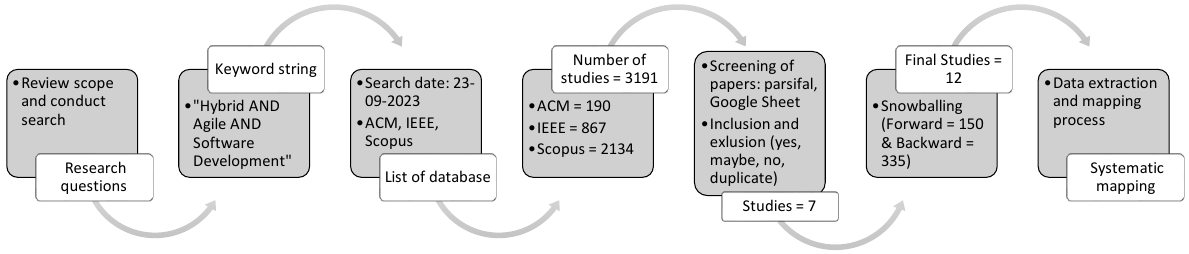}
    \caption{Research Steps based on guidelines mentioned in the study \cite{petersen2008systematic, petersen2015guidelines}},
    \label{fig: Research steps}
\end{figure*}

Figure \ref{fig: Research steps} shows the steps we followed in conducting the systematic mapping study, according to the guidelines by Petersen et al. \cite{petersen2008systematic, petersen2015guidelines}. 

\subsection{Search strategy}

\subsubsection{Keyword search} \label{ks}
We used the AND logic between the central concepts, i.e., "Hybrid" AND "Agile" AND "Software development". We used OR logic between the synonyms of sets 1, 2 and 3. Later, we combined the three sets to reach the final search string, i.e., \textbf{({SET 1} AND {SET 2} AND {SET 3})}. Following are the three sets:
\begin{itemize}
    \item SET 1: Scoping the search for hybrid work. Therefore, we consider \textbf{\textit{(hybrid work OR blended work OR Covid-19 OR distributed teams OR mixed work OR online collaboration OR pandemic OR remote and on-site work OR remote collaboration OR virtual collaboration OR virtual teams OR work from anywhere OR work-from-home OR e-work OR flexible OR telecommunication OR telework OR teleworking)}}.
    \item SET 2: Search terms are directly related to agile. Hence, we consider \textbf{\textit{(XP OR agile OR agile programming OR extreme programming OR Kanban OR lean OR scaled agile OR scrum)}}.
    \item SET 3: Search term is related to software development. Hence, we consider \textbf{\textit{(SE teams OR programming team OR programming teams OR software developer OR software developers OR software development OR software engineer OR software engineering OR software evolution OR software life cycle OR software maintenance OR software process OR software professional OR software professionals OR software project OR software projects OR software quality OR software system OR software systems OR software team OR software teams)}}.
    \end{itemize}

\subsection{Screening and selection}
We used three well-known academic databases to retrieve the studies \cite{kitchenham2009impact}. The three libraries we used are \href{https://dl.acm.org}{ACM Digital Library}, \href{https://ieeexplore.ieee.org/Xplore/home.jsp}{IEEE Xplore}, and \href{https://www.scopus.com/search/form.uri?display=basic#basic}{Scopus}. 
We conducted our search on 23-09-2023, as keywords, abstract and title search. As shown in Figure \ref{fig: Research steps}, we obtained 3191  studies (ACM: 190, IEEE: 867, and Scopus: 2134). We imported all studies into \href{https://parsif.al}{Parsifal}, an online tool that helps researchers conduct systematic reviews within software engineering.  
Next, we used the following inclusion and exclusion criteria to select the studies. An included study had to fulfill all the inclusion criteria, whereas for exclusion, fitting to one or more criteria would exclude the paper. 

 \begin{table*}[t]
\caption{The list of the reviewed studies}
\label{tab:ele}
\vspace{-5pt}
\centering
\footnotesize
\begin{tabular}{ l p{15.8cm} l } 
 \toprule
 \multicolumn{1}{l}{\textbf{No.}} &
\multicolumn{1}{l}{\textbf{Authors and \textit{title of the study}}} &
\multicolumn{1}{l}{\textbf{Year}} \\ [0.5ex] 
 \midrule
S1 & \textbf{Advait Deshpande, Helen Sharp, Leonor Barroca, and Peggy Gregory} \textit{"Remote Working and Collaboration in Agile Teams"}\nociteS{deshpande2016remote} & 2016 \\ \addlinespace
S2 &  \textbf{Pernille Lous, Paolo Tell, Christian Bo Michelsen, Yvonne Dittrich, and Allan Ebdrup} \textit{"From Scrum to Agile: a Journey to Tackle the Challenges of Distributed Development in an Agile Team"} \nociteS{lous2018scrum} & 2018 \\ \addlinespace
S3 & \textbf{Michael Neumann, Habibpour Daryosch, Eichhorn Dennis, John A, Steinmann
Stefan, Farajian L, and David, Mötefindt} \textit{"What Remains from Covid-19? Agile Software Development in Hybrid Work Organization: A Single Case Study"} \nociteS{neumann2022remains} &  2022 \\ \addlinespace
S4 & \textbf{Darja Smite, Nils Brede Moe, Anastasiia Tkalich, Geir Kjetil Hanssen, Kristina Nydal, Jenny Nøkleberg Sandbæk, Hedda Wasskog Aamo, Ada Olsdatter Hagaseth, Scott Aleksander Bekke, and Malin Holte} \textit{"Half-Empty Offices in Flexible Work Arrangements: Why are Employees Not Returning?"} \nociteS{smite2022half} & 2022 \\ \addlinespace
S5 & \textbf{Tor Sporsem and Nils Brede Moe} \textit{"Coordination Strategies When Working from Anywhere: A Case Study of Two Agile Teams"} \nociteS{sporsem2022coordination} & 2022 \\ \addlinespace
S6 & \textbf{Tor Sporsem, Audun Fauchald Strand, and Geir Kjetil Hanssen} \textit{"Unscheduled Meetings in Hybrid Work"} \nociteS{sporsem2022unscheduled}  & 2022 \\ \addlinespace
S7 & \textbf{Anastasiia Tkalich, Darja Smite, Nina Haugland Andersen, and Nils Brede Moe} \textit{"What Happens to Psychological Safety When Going Remote?"} \nociteS{tkalich2022happens} & 2022 \\ \addlinespace
S8 & \textbf{Zhendong Wang, Yi-Hung Chou, Kayla Fathi, Tobias Schimmer, Peter Colligan, David Redmiles, and Rafael Prikladnicki} \textit{"Co-designing for a Hybrid Workplace Experience in Software Development"} \nociteS{wang2022co} & 2022 \\ \addlinespace
S9 &  \textbf{Jedrzej Bablo, Bartosz Marcinkowski, and Adam Przybylek} \textit{"Overcoming Challenges of Virtual Scrum Teams: Lessons Learned Through an Action Research Study"} \nociteS{bablo2023overcoming} & 2023 \\ \addlinespace
S10 & \textbf{Safinaz Buyukguzel and Ufuk Balaman} \textit{"The spatial organization of hybrid Scrum meetings: A multimodal conversation analysis study"} \nociteS{buyukguzel2023spatial} & 2023 \\ \addlinespace
S11 & \textbf{Safinaz Buyukguzel and Robb Mitchell} \textit{"Progressivity in Hybrid Meetings: Daily Scrum as an Enabling Constraint for a Multi-Locational Software Development Team"} \nociteS{buyukguzel2023progressivity} & 2023 \\ \addlinespace
S12 & \textbf{Kai-Kristian Kemell and Matti Saarikallio} \textit{"Hybrid Work Practices and Strategies in Software Engineering-Emerging Software Developer Experiences"} \nociteS{kemell2023hybrid} & 2023 \\ \addlinespace
 \bottomrule
\end{tabular}
\end{table*}

\subsubsection{Inclusion Criteria}
 \begin{enumerate}
     \item Studies conducted in professional settings in hybrid and agile environments.
    \item Studies published in the time frame since the establishment of the Agile Manifesto \cite{beck2001agile}, hence, from 2001 to Sept 23rd, 2023.
    \item Studies that provide an answer to at least one of our research questions.
 \end{enumerate}
     
\subsubsection{Exclusion Criteria}
 \begin{enumerate}
     \item Studies outside the scope of hybrid work and agile software development.
    \item Editorials, invited papers, research proposals, self-archived, non peer-reviewed studies,  non peer-reviewed books or book chapters, summaries of conference proceedings with titles, theses and dissertations, reports, technical reports, working papers, white papers, patents, or grey literature (magazines, articles).
    \item Studies that are duplicates of other studies.
    \item Studies outside the professional work settings, e.g., student or education projects and secondary studies.
     \item Studies that are not written in English.
 \end{enumerate}
 
 \subsubsection{Title, abstract, full text screening, and data extraction}
 Two authors reviewed the titles and abstracts of all 3191 papers. From \href{https://parsif.al}{Parsifal}, then we exported the results to Google Sheets. We categorised each paper into \textit{`accept or yes', `maybe', `reject or no' and `duplicate'}. In the screening process we found 756 \textit{`duplicate'} and 2409 \textit{`reject or no'} studies. The 21 studies categorised \textit{`maybe'} were discussed by four authors and based on the discussion we classified either to \textit{`accept'} or \textit{`reject'} the paper. We found seven papers from the main search that passed the inclusion and exclusion criteria.

Subsequently, we performed forward and backward snowballing on these seven papers, as well as a few key papers (e.g., \cite{conboy2023future}). We found 150 candidate publications from forward and 335 from backward snowballing for further screening. After using the inclusion and exclusion criteria, we found five relevant papers. The papers we obtained from the forward snowballing  search are: S4, S6, S7, S8, S12 (see table \ref{tab:ele}). We obtained no studies from the backward snowballing search. In total, we performed full-text screening for: 7 (main search) + 5 (forward snowballing) = 12 primary studies for conducting systematic mapping. We developed a data extraction sheet, in order to collect information from these studies. The data extraction form can be retrieved online\footnote{https://doi.org/10.6084/m9.figshare.24547204.v1}.

In the next section, we will provide answers to the research questions asked in this mapping study.

\section{Results}
\subsection{RQ1.) What are the publication trends and characteristics of existing research on hybrid work in agile software development?}
Table \ref{tab:ele} shows the list of the reviewed studies sorted by year with authors' names and titles of the papers. In the table's first column, we mentioned the serial number, i.e., S1--S12. The second column first describes the author's names and then the study's title. Finally, the last column outlines the year the study was published. We can notice in Table \ref{tab:ele} that several authors had two or more publications, for example, (S5 \& S6), (S5 \& S7), (S4 \& S7), and (S10 \& S11).

\subsubsection*{RQ1.1.) What are the publication years and types of research articles on hybrid work in agile software development?}
\begin{figure}[H]
    \centering
    \includegraphics[width=0.9\linewidth]{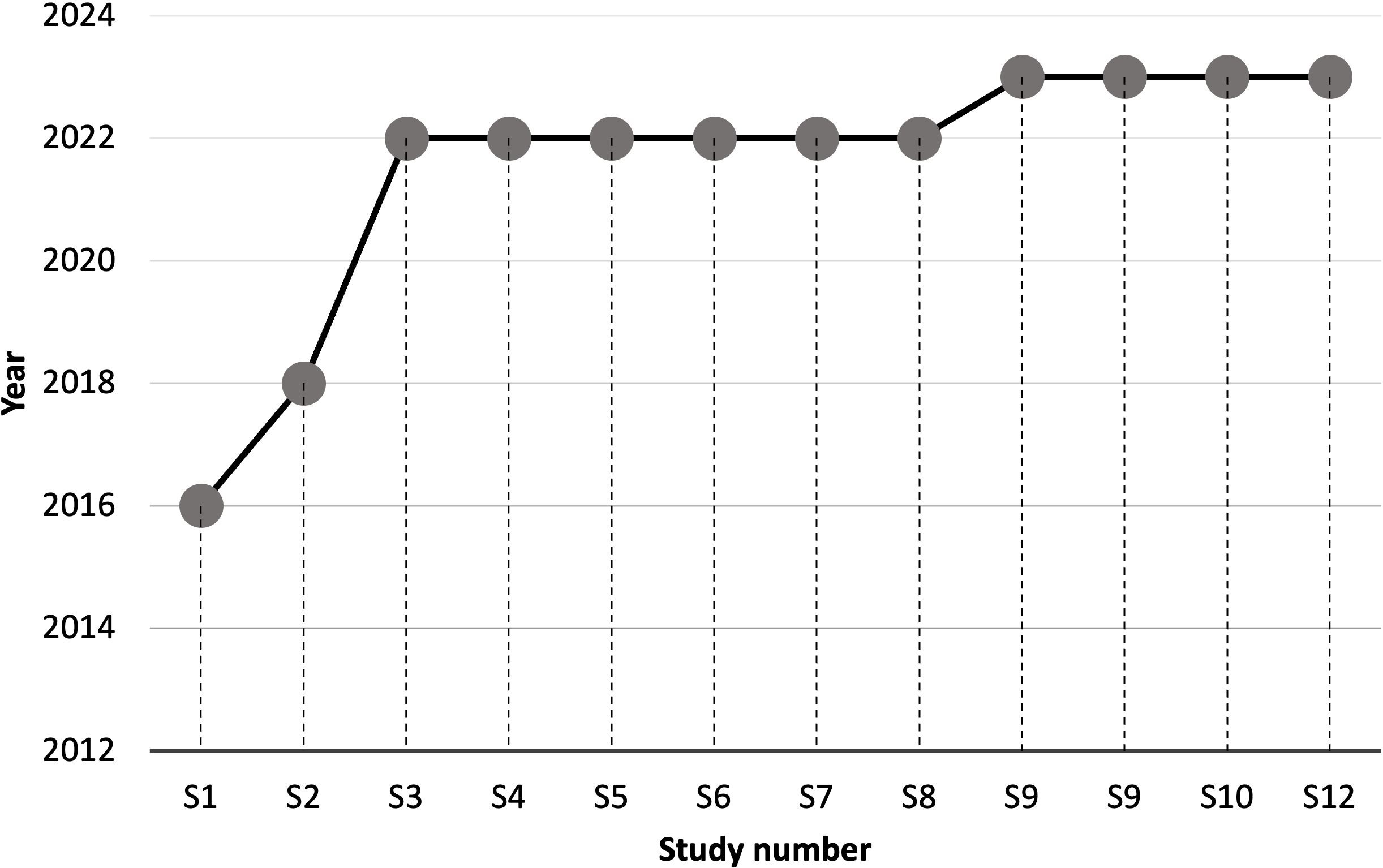}
    \caption{Yearly distribution of the studies}
    \label{fig:year1}
\end{figure}

\begin{figure}[H]
    \centering
    \includegraphics[width=0.6\linewidth]{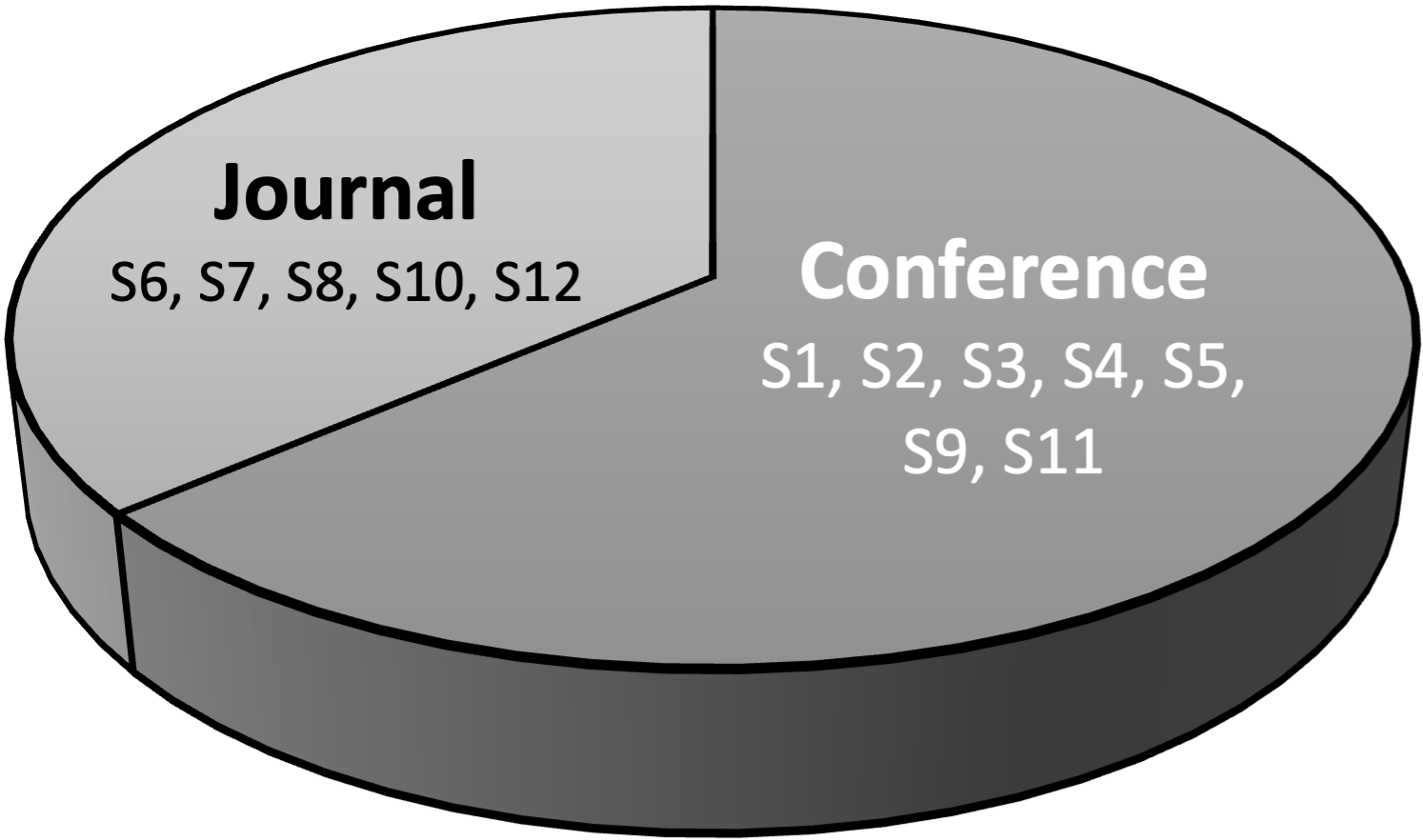}
    \caption{Publication types: journals vs. conferences}
    \label{fig:jour}
\end{figure}

As evident from Figure \ref{fig:year1}, we can see a rising trend in the number of papers after Covid-19. The horizontal axis in Figure \ref{fig:year1} represents the study number, and the vertical axis represents years. Only two papers (S1, S2) were published prior to the pandemic, i.e., in 2016 and 2018 respectively. The remaining studies (S3--S12) were published in 2022 and 2023. As visible in Figure \ref{fig:jour}, seven primary studies were published in conferences, while five were published in journals. Studies S6, S7 and S8 were published in IEEE Software, and studies S10 and S12 were published in Discourse \& Communication, and IEEE Access respectively. 

The conferences were: International Conference on Information Systems (S1), International Conference on Software and System Process (S2),  International Conference on Software Engineering Research and Innovation (S3), International Conference on Product-Focused Software Process Improvement (S4), International Conference on Agile Software Development (S5, S9), and ACM Conference On Computer-Supported Cooperative Work And Social Computing (S11). Thus, only one conference had published two papers that were included.

\subsubsection*{RQ1.2.) Which research methods have been employed in the published studies for hybrid work in agile software development?}

\begin{figure}[H]
    \centering
    \includegraphics[width=0.6\linewidth]{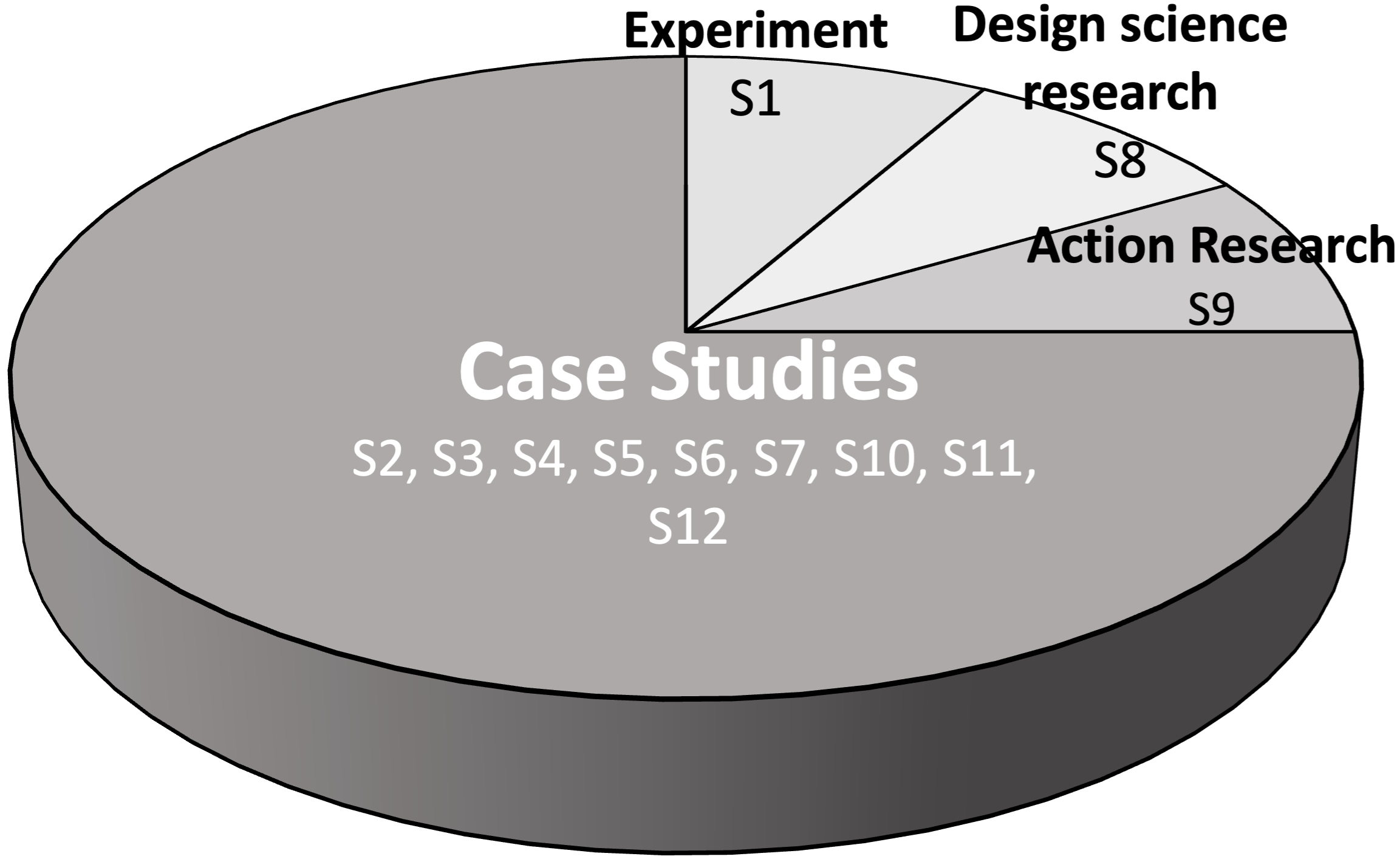}
    \caption{Types of empirical studies}
    \label{fig:emp}
\end{figure}

All twelve primary studies applied empirical research methods (see Figure \ref{fig:emp}). We did not find any conceptual studies, position papers, systematic literature reviews, systematic mapping studies, or multivocal literature reviews. From the obtained studies, we found that the  majority used case study as the empirical method to conduct the investigation. This method was applied in nine studies (S2--S7, S10--S12). The remaining three studies used experiment (S1), design science research (S8), and action research (S9).

\begin{figure*}[t]
    \centering
    \includegraphics[width=0.9\linewidth]{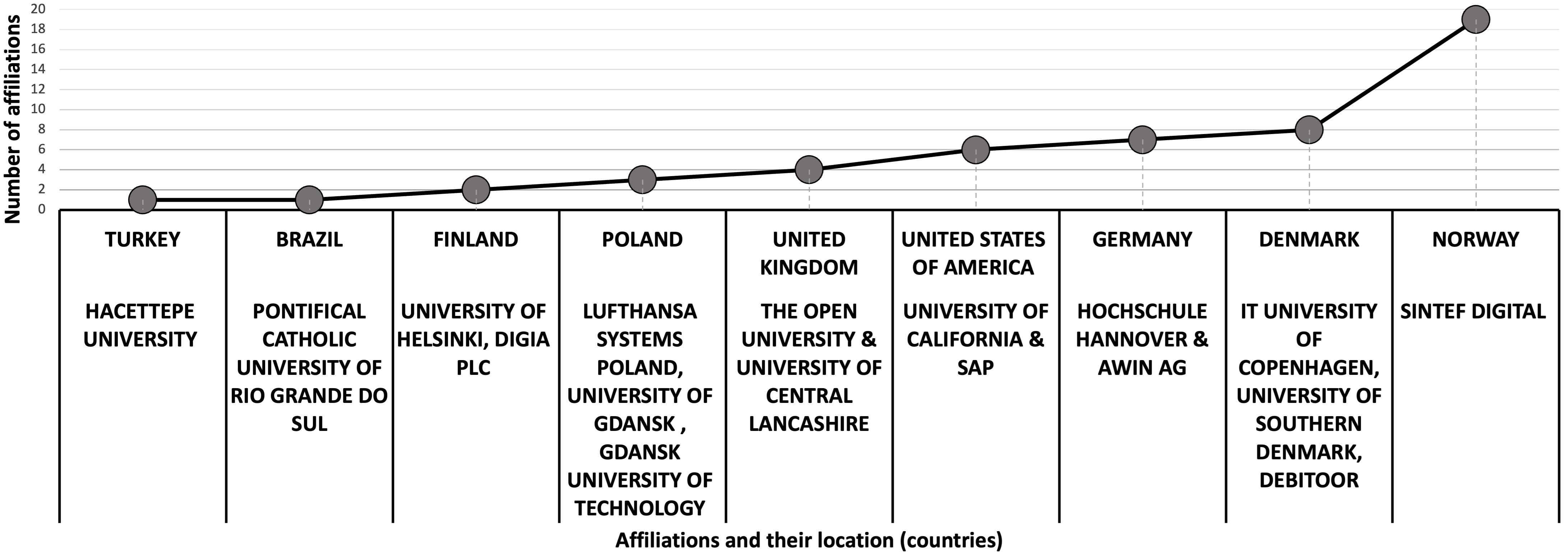}
    \caption{Authors affiliations and their locations}
    \label{fig:afil}
\end{figure*}

\begin{figure*}[t]
    \centering
    \includegraphics[width=0.9\linewidth]{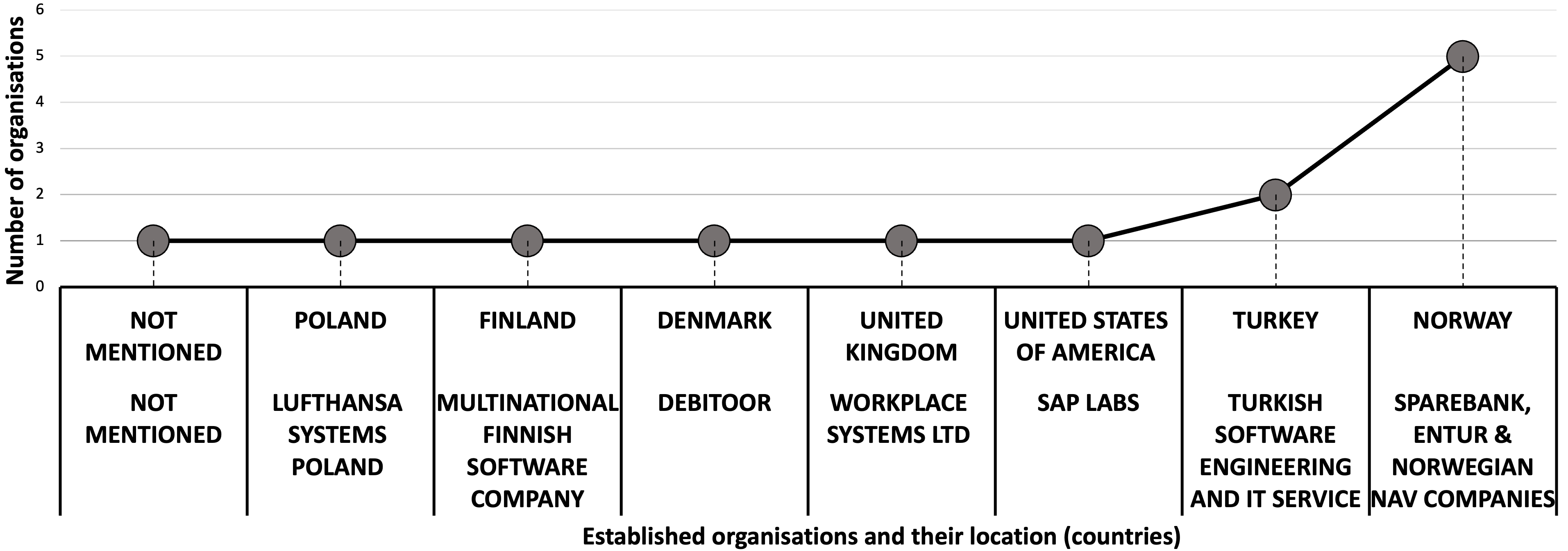}
    \caption{The studied organisations and their locations}
    \label{fig:orc}
\end{figure*}

The data collection techniques used in the primary studies included interviews (S1, S2, S4, S5, S7--S9, S12), observations (S1, S2, S5, S6, S9--S11), surveys (S3, S4, S7, S9), group discussion (S1), notes from virtual stand-ups (S5),  collection of other available data, i.e., office attendance, meetings data, and strategy documents (S7), co-design workshops (S8), and focus groups (S9).

It is noticeable that most of these studies (ten, S1--S3, S5--S11) have `team' as the level of analysis,  whereas S4 and S12 used `individual'. 

\subsubsection*{RQ1.3.) In which countries and organisations has the research been carried out for hybrid work in agile software development?}

We found that the research has primarily been carried out in European countries, with Norway dominating the research trend, as visible in both Figures \ref{fig:afil} and \ref{fig:orc}. Only one study, S8, was carried out in the USA (California).

The horizontal axis of Figure \ref{fig:afil} represents the country and affiliation of the authors, while the vertical axis shows the count of the affiliation. For example, Figure \ref{fig:afil} shows that 19 authors are from Norway, all affiliated with `SINTEF Digital'. Whereas, a total of eight authors are located in Denmark, under three different affiliations: IT University of Copenhagen (4 authors), University of Southern Denmark (3 authors), and Debitoor (1 author). The other countries represented in the studies are Poland, Finland, United Kingdom, Turkey, USA, and Brazil. Focusing on the affiliation, the highest number of researchers (19) are from `SINTEF Digital, Norway', followed by eight research affiliations from Denmark including `IT University of Copenhagen, University of Southern Denmark, and Debitoor, Denmark', and seven research affiliations from Germany, including `Hochschule Hannover', and `AWIN AG'.

 \begin{table*}[ht]
\caption{Research questions mapped to \textit{the conceptual framework for organizing research
questions on hybrid work in SE} \cite{paas}}
\label{tab:ss}
\vspace{-5pt}
\centering
\footnotesize
\begin{tabular}{ l p{8.2cm} l p{2.8cm} l} 
 \toprule
\multicolumn{1}{l}{\textbf{No.}} &
\multicolumn{1}{l}{\textbf{Research questions}} &
\multicolumn{1}{l}{\textbf{Factors}} &
\multicolumn{1}{l}{\textbf{Hybrid Work in SE}} &
\multicolumn{1}{l}{\textbf{Impact}} \\ [0.4ex] 
 \midrule
S1 & RQ1: What kind of issues (if any) are faced by a hybrid agile team in practice?  &  &  & Process  \\
& RQ2: How does a hybrid agile team with a remote worker collaborate?  & & Process, People (Individual) &  \\
& RQ3: How does the experience of remote working compare between a remote worker and his colocated teammates in a hybrid agile team? & & &  Process \\
S2 & RQ: How can a development team adopt an agile process without compromising on the core values and principles when facing the challenges of distributed development? & &Process  & \\
S3 & RQ1: Which work organization type (remote or onsite) is preferred by the members of agile software development teams operating currently in a remote work setting?  &  Process, People (Individual) & Process, People (Individual) &  \\
& RQ2: Which organization type (remote or onsite) support the use of specific agile practices from a team member point of view of agile software development teams?  &   & Process, People (Individual) &  \\
&  RQ3: How can we perform agile practices in a hybrid work setting?&  & Process & \\
S4 &RQ1: How often are employees present in the office? & People (Individual) &  &  \\
& RQ2: What hinders and what motivates employees to visit the office?& People (Individual) & &  \\
S5 & RQ: What coordination strategies are used by agile teams when working from anywhere?& Process & & \\
S6 & RQ: How do developers maintain unscheduled meetings in post-lockdown work life? &  & Process, People (Individual)& \\    
S7 & Research objective: The study investigates how different work modes affect software teams’ psychological safety. &  & & People (Team)\\
S8 & RQ: How to design for and continuously improve a workplace experience with software development teams while transitioning to a hybrid work arrangement? &  & Process, People (Team and Organisation) & \\
S9 &   RQ1: How did Scrum teams adapt their practices and processes due to the ad-hoc shift to remote work?&  & Process & \\
&  RQ2: What are the advantages of remote work for Scrum team members?  &  &  & Process \\
&  RQ3: What new challenges are faced by virtual Scrum teams and their members?  &  &  &  Process\\
& RQ4: How can these challenges be mitigated?  &  &  & Process  \\
S10 & Research objective: The study examined hybrid Scrum meetings in order to understand how space is co-constructed in them. & & Process & \\
S11 & RQ: How do participants ensure progressivity and get the work done in hybrid meetings despite disruptions? &  & Process & \\
S12 & RQ1: What factors influence the work mode preferences of software developers, and why?  & People (Individual) & & \\
& RQ2: How could alternating between remote work and working on-site alleviate the challenges associated with each work mode while leveraging the benefits of both?  & & & People (Individual)\\
[1ex] 
 \bottomrule
\end{tabular}
\end{table*}

Concerning the studied organisations, the horizontal axis in Figure \ref{fig:orc} represents the organisations that are studied in the articles, and the country they are located in. The vertical axis shows the count of the organisations which are represented. For example, in Figure \ref{fig:orc}, the count is five for Norway. This is because Sparebank was involved in two studies (S4, S7), and Entur in two (S5, S6), and Norwegian Labour and Welfare Administration (NAV) was involved once (S6). Whereas for Turkey, a Turkish software engineering \& IT service providing company was involved in both S10 and S11, summing up the count to two. As shown in Figure \ref{fig:orc} one study (S3) did not mention the organisation and country information. But referring to the previous Figure \ref{fig:afil}, the author's affiliation belongs to `Hochschule Hannover, and AWIN AG, Germany'. 

\subsection{RQ2.) Which research questions have been investigated in hybrid work in agile software development?}

As shown in Table \ref{tab:ss}, a total of 19 research questions were investigated in the reviewed studies: 11 (How), 6 (What), and 2 (Which). In studies S7 and S10 only the research objective was stated, which is included in Table \ref{tab:ss}. We mapped the research questions and objectives according to \textit{the conceptual framework for organizing research questions on hybrid work in SE} \cite{paas}, to show both the categories and perspectives represented. 
We found that the largest number of research questions (10) could be mapped to the \textit{hybrid work in SE} category (S1--S3, S6, S8--S11). Following this, seven research questions were mapped to the \textit{impact} of hybrid work (S1, S7, S9, S12), and five research questions to influencing \textit{factors} (S3--S5, S12). 

As can be seen in Table \ref{tab:ss}, only one of the research questions was mapped to more than one category (S3: RQ1). However, multiple research questions could be mapped to more than one of the ``3 P's'' perspectives of the framework \cite{paas}. Mostly, the studied research questions fall under the \textit{process} perspective, followed by the \textit{people (individual)} perspective. Just two map to the \textit{people (team)} perspective, while one mapped to the \textit{people (organisation)} perspective. Our mapping did not reveal any research questions in these studies that could be mapped to the \textit{product} perspective.

\subsection{RQ3.) In which kind of hybrid settings is agile software development carried out?}
The following work arrangements, online tools for hybrid work, hybrid work policies, agile frameworks, practices, and roles have been identified in the reviewed studies.

\subsubsection*{Work arrangements:} \label{wa}Based on the team typology and model presented in Figure \ref{fig:wa} from  \cite{smite2022future}, we identified the following work arrangements, which are shown in Table \ref{tab:hybset}. As previously mentioned, in contrast to all the other primary studies, the level of analysis in S4 and S12 was carried out on a individual basis, so team work arrangements were not investigated. 

Regarding the work location arrangements, \textbf{partially aligned} teams were identified in five of the studies, with two of these studies specifying \textit{semi-remote mode} (S1, S2), while the remaining three specified \textit{office-remote mix mode} (S5--S7). \textbf{Variegated} teams were identified in three of the studies as well, with two of the studies specifying work in \textit{aligned office-remote mix mode} (S6, S8) and the other in \textit{aligned remote-first mode} (S9). In addition, \textbf{hybrid} teams working in \textit{flexible location mode} were identified in S7. The remaining studies (S3, S4, S10--S12) do not map to exact arrangements as described in \cite{smite2022future}. 

Regarding the work schedule arrangements, teams in studies S6, S7, and S9 worked in \textit{synchronous schedule mode} on the planned office days, but the schedule of the remaining work days was not specified. Teams in study S8 also worked in \textit{synchronous schedule mode} on the planned office days, but worked in \textit{core meeting mode} and \textit{core hours mode} during planned WFX days. \textit{Core meeting mode} was also identified in studies S1, and S7. The work schedule was not described for one of the teams in studies S6 and S7, or in the remaining seven studies (S2--S5, S10--S12). 

\begin{table}[h]
\caption{Work arrangements mapped to team typology and model \cite{smite2022future}}
\label{tab:hybset}
\vspace{-4pt}
\centering
\footnotesize
\begin{tabular}{ l p{3.5cm} p{3.5cm} } 
 \toprule
 \multicolumn{1}{l}{\textbf{No.}} &
\multicolumn{1}{l}{\textbf{Work location modes}} &
\multicolumn{1}{l}{\textbf{Work schedule modes}} \\ [0.3ex] 
 \midrule
S1	& \textit{Semi-remote} (partially aligned team)  &\textit{Core meeting} \\
S2	& \textit{Semi-remote} (partially aligned team)  & Not specified \\
S3	& Alignment not specified & Not specified \\
S4	& Not specified & Not specified \\
S5	& \textit{Office-remote mix} (partially aligned team) & Not specified \\
S6	& Team 1: \textit{Aligned office-remote mix} (variegated team) | Team 2: \textit{Office-remote mix} (partially aligned team) & Team 1: \textit{Synchronous schedule} on planned office days (other days not specified) | Team 2: Not specified \\
S7	& Team 1 \& 3: \textit{Flexible location} (hybrid team) | Team 2: \textit{Office-remote mix} (partially aligned team) & Team 1: \textit{Core meeting} | Team 2: \textit{Synchronous schedule} on planned office days (other days not specified) | Team 3: Not specified \\
S8	&  \textit{Aligned office-remote mix} (variegated team) & \textit{Synchronous schedule} on planned office days, \textit{core hours \& meeting} other days \\
S9	& \textit{Aligned remote-first} (variegated team) & \textit{Synchronous schedule} on planned office days (other days not specified) \\ 
S10	& Alignment not specified & Not specified \\
S11	& Alignment not specified & Not specified \\
S12	& Not specified & Not specified \\ 
 \bottomrule
\end{tabular}
\end{table}

\subsubsection*{Online tools:} 
 In total, 35 online tools used in hybrid work to support agile software development were identified in the reviewed studies. Only one of the studies (S4) did not mention any online tools. The number of mentions for each tool can be viewed in Table \ref{tab:tools}. Atlassian Jira and Slack were both mentioned in six studies. Microsoft Teams was mentioned in five studies, while Atlassian Confluence, digital whiteboards, and Mural were mentioned in three studies. Discord, Microsoft Outlook, and Miro were mentioned in two studies, and the remaining 26 tools were mentioned only once.

 \begin{table}[h]
\caption{ Online tools to support hybrid work in agile software development}
\label{tab:tools}
\vspace{-4pt}
\centering
\footnotesize
\begin{tabular}{ p{6.7cm} c} 
 \toprule
\multicolumn{1}{l}{\textbf{Online tools}} &
\multicolumn{1}{l}{\textbf{Mentions}} \\ [0.3ex] 
 \midrule
Atlassian Jira, Slack & 6 \\ 
Microsoft Teams & 5 \\ 
Atlassian Confluence, Digital Whiteboards, Mural & 3 \\ 
Discord, Microsoft Outlook, Miro & 2 \\ 
Digital Boards (Kanban, Scrum), Appear.in, Asana, Atlassian Bitbucket, GitHub, Google Hangouts, Google Sheets, Google Slides, Hatjitsu, Hipchat, Lucidchart, Mentimeter, Microsoft SharePoint, MySQL,  PhpStorm, PlanITPoker, Scrum Poker, Skype for Business, Sourcetree, TeamRetro, Timbo, Trello, Waffle.io, Webex, Zoom & 1 \\  
 \bottomrule
\end{tabular}
\end{table}

\subsubsection*{Hybrid work policies:} In Table \ref{tab:poli} we present the hybrid work policies identified in the studies, mapped to the \textit{people} perspectives of the framework (Figure \ref{fig:3p}) \cite{paas}, and the underlying \textit{individual}, \textit{team}, and \textit{organisational} levels. Mostly, the studies show that the company policies around hybrid work can be mapped to the team (S1, S5, S6, S8) or individual (S2--S4) levels, as the employees or teams are allowed to work \textbf{wherever}. In addition, S3 also states that work can be carried out \textbf{whenever}. The hybrid work policy in S7 can be mapped to both the individual and team levels, allowing work to be carried out \textbf{wherever}. 

Only three studies revealed policies that could be mapped to the organisational level, with S9 requiring employees to be present at the office one day per week, and S12 required only an occasional day at the office. The policy in S8 specified three mandatory office days per week, and a flexible schedule for the two WFX days (outside of core hours and meetings). The remaining two reviewed studies, S10 and S11, did not reveal clear policies for hybrid work. 

\begin{table}[h]
\caption{Hybrid work policy mapped to the \textit{conceptual framework for organizing research questions on hybrid work in SE} \cite{paas}}
\label{tab:poli}
\vspace{-4pt}
\centering
\footnotesize
\begin{tabular}{ p{1cm} p{3.1cm} p{3.2cm} } 
 \toprule
 \multicolumn{1}{l}{\textbf{No.}} &
\multicolumn{1}{l}{\textbf{People perspective}} &
\multicolumn{1}{l}{\textbf{Hybrid work policy}} \\ [0.3ex] 
 \midrule
S1	& Team & Wherever \\
S2	& Individual & Wherever \\
S3	& Individual & Wherever \& whenever \\
S4	& Individual & Wherever \\
S5	& Team & Wherever \\
S6 & Team & Wherever \\
S7	&Individual \& team	& Wherever \\
S8 & Team  \& organisation &  Three mandatory office days, flexible schedule during WFX (outside of core hours and meetings)\\
S9	&Organisation&	One mandatory office day \\ 
S10	&Not specified&	Not specified \\
S11	&Not specified&	Not specified \\
S12	&Organisation	&Occasional office day \\ 
 \bottomrule
\end{tabular}
\end{table}

\subsubsection*{Agile frameworks, practices, and roles:} Table \ref{tab:agile} outlines the entire list of agile frameworks, practices, and roles, which were mentioned in the studies. We found that seven of the primary studies (S1--S3, S8--S11) described Scrum specifically as the agile framework used, while three of the studies discussed Kanban (S2, S4, S12).  The remaining studies mentioned varying agile frameworks, practices, and roles, including Scrum Master (SM), Product Owner (PO), product manager (PM), Sprint Retrospective, Sprint Planning, backlog meetings, product backlog refinement meetings, and daily stand-ups. Study S8 provided the most detailed account of agile practices, as can be seen in Table \ref{tab:agile}. Studies S6 and S7 did not mention any specific agile frameworks, practices or roles, but the companies involved were the same as in S5 and S4, respectively.

\begin{table} [H]
\caption{Agile frameworks, practices, and roles}
\label{tab:agile}
\vspace{-4pt}
\centering
\footnotesize
\begin{tabular}{ l p{1.8cm} p{3.6cm} p{1.1cm}}
 \toprule
 \multicolumn{1}{l}{\textbf{No.}} &
\multicolumn{1}{l}{\textbf{Agile frameworks}} &
\multicolumn{1}{l}{\textbf{Practices}} &
\multicolumn{1}{l}{\textbf{Roles}} \\ 
 \midrule
S1	& Scrum & Agile release cycle, daily stand-ups & SM, PO, tester  \\
S2	& Scrum, Kanban & & \\
S3	& Scrum  & Sprint, Scrum poker & Quality specialist, SM \\
S4	& Scrum, Kanban & Backlog meetings, daily stand-ups &  \\
S5	& Various agile methods & Virtual stand-ups &  \\
S6  & Not specified & &  \\
S7	& Not specified &	&  \\
S8	& Large-Scale Scrum (LeSS) & Sprint Planning, Sprint Review, team retrospective, overall monthy retrospective, multi-team review, daily stand-up, SM sync, overall PBR, single-team PBR, multi-team PBR, collaborative sessions (software and UX design, peer code reviews, pairing) & Software engineers, UX designers, PM \\
S9	& Not specified & Scrum Events, Sprint Planning, team retrospective, poker & \\ 
S10	& Scrum & & \\
S11	& Scrum & Daily Scrum & \\
S12	& Kanban & & SM \\ 
 \bottomrule
\end{tabular}
\end{table}

\section{Discussion}
The findings from our mapping study show that the research at the intersection between hybrid work and agile software development is on the rise, especially after the Covid-19 pandemic, with six of the reviewed studies published in 2022, and four in 2023 (by September 23rd). It is expected that the trend will keep increasing as more software companies speculate and experiment on hybrid work, the ``new normal'' way of working \cite{paas}. 

With the rise of published articles \cite{wang2023many}, the understanding of hybrid work concerning work settings may be enhanced and strengthened. The small number of selected studies, only twelve papers, was lower than what we expected. The reason for that might be that hybrid work was not a big thing before the Covid-19 pandemic and only the pandemic gave rise to its' popularity. Studies performed during the Covid-19 pandemic were done mainly in fully remote settings (e.g., \cite{ralph_pandemic_2020,russo_predictors_2021}), instead of hybrid, due to the lockdown of society. Thus, it seems that after the pandemic only a few studies have been published so far, most probably due to research lagging a bit behind, since it takes time for researchers to conceive, design, implement and eventually publish a piece of research. However, companies are performing their own internal studies, as they are desperately trying to find solutions to the current hybrid work situation (e.g. Spotify\footnote{https://newsroom.spotify.com/2021-02-12/distributed-first-is-the-future-of-work-at-spotify/}). Thus, our main finding is that this research topic, hybrid work in agile software development, is surprisingly under-researched and would need more empirical studies so that research could support software engineering companies that are currently pondering what to do and how to best organize hybrid work.

Our second, interesting finding was the fact that the studies found were mainly performed in Europe and studied European organizations, while only one paper (S8) came from outside Europe, from the US. Moreover, seven papers came from the Nordic countries, with five from Norway, one from Denmark, and one from Finland. Thus, it seems that this topic is interesting especially for European companies and researchers and even more so in the Nordic countries. The reason for this might be that there are stronger mandates for knowledge workers in other parts of the world to return to their offices than their European counterparts, so hybrid work is not such a big thing there. In addition, especially Nordic software engineers seem to have a stronger preference for hybrid work. However, 
hybrid work is a global phenomenon, and to better understand whether it is the ``new normal'' way of working and is here to stay for a long time, 
we encourage more research on hybrid work in agile organisations in other geographical locations.

Covid-19 forced software engineers to move to using fully digital tools and replace whiteboards, post-it notes and markers with online tools. Many of the same tools are still in use while in the hybrid setting. In total, our study revealed 35 different online tools used to support hybrid work in agile software development. In the future, it would be interesting to study whether these online tools provide adequate support for hybrid work, or if new and different types of supporting tools are needed, such as physical tools to boost innovation. It would be interesting to study as well how agile software teams are using these tools and how they could best support hybrid work.

The analysis of the research questions asked in the reviewed studies (answer to RQ2, Table \ref{tab:ss}) shows a strong concentration of the \textit{process} perspective on hybrid work taken in these studies, and much less studies taking the \textit{people} perspective. Given the fact that the reviewed studies are conducted in organisations that apply agile methods, and agile methods embrace ``focus on people'' as a key agile value, we found this finding intriguing. It indicates a potential research gap that deserves the attention of more researchers active in the intersection of hybrid work and agile software development research areas. For example, future research could be carried out taking the \textit{people} perspective at the organisational level, to investigate how organisational factors could influence hybrid work settings, or how agile organisations could better arrange hybrid work, and/or what are the consequences of such work arrangements on the innovation capability of the organisation. 

As evident in Section \ref{wa}, the mapping of hybrid work arrangements to the model presented in \cite{smite2022future} shows that over half of the primary studies (S2, S5--S7, S10, S11) do not investigate the work schedule of all the software teams involved. Similarly, the schedule of one team from both studies S6 and S7, and the teams in S9, is only apparent during the planned office days, where the teams work together onsite. In addition, results from the work location mapping show that the alignment of teams is not always investigated. We recommend that future studies could include the varying work schedule modes of hybrid work, as well as the degree of team alignment regarding the work location, as mentioned in the model \cite{smite2022future}.

Since we are mapping the research terrain at the intersection of hybrid work and agile software development, an interesting topic to ponder on when conducting a study in this area is, which one is the focus of the study, `hybrid work' or `agile software development'? The emphasis of `hybrid work' or `agile software development' changes the perspective of the study. Most reviewed studies focused on hybrid work, taking agile software development as research settings. Relatively less studies have investigated the adoption and adaptation of agile methods in hybrid work settings. More studies along this line would be a welcome addition to the body of knowledge of agile research. 
Emphasizing `agile software development' before `hybrid work' could reveal specific hybrid work settings that are just supporting or beneficial for executing agile software development. 

\paragraph{Threats to validity}
This study might have a selection bias where the specific criteria we considered to include and exclude the study might bring biases to the results. To overcome this, we carefully followed the guidelines mentioned in \cite{petersen2008systematic} and \cite{petersen2015guidelines}. Four of the five authors were on board to finalise the inclusion and exclusion criteria. Another threat could be the search strategy bias with the keywords. To overcome this, we considered several synonyms of the main three keywords and formulated them into sets as shown in Section \ref{ks}. We conducted pilot studies with various combinations of the sets, and lastly, we finalised the mentioned keywords. Finally, as the search was conducted on title, keywords and abstract fields of the articles, and not to the full text, some papers not mentioning our search terms in those parts, could have been omitted. To mitigate this, we did forward and backward snowballing with the selected papers and a few key papers.

The study could face threats related to data extraction, which could lead to several errors and inconsistencies. To minimise this threat, four authors were involved during the data extraction procedure. In addition, two authors reviewed all the initial 3191 studies and 485 snowballing studies. 

\paragraph{Limitation of the study}
The number of the reviewed studies presents the major limitation, as we obtained results from only 12 studies. This limited literature collection constrained the breadth of our systematic mapping study. Another limitation is the variation of keywords, i.e., the use of `hybrid' or `agile', which was previously discussed. In line with this limitation is the use of keywords, as synonyms of software development could change the number of results in the study.

\section{Conclusions}
This systematic mapping study provides a broad overview of the emergent research area at the intersection of hybrid work and agile software development. We screened 3191 papers obtained through the keyword search and 485 through snowballing. After applying inclusion and exclusion criteria, we selected 12 primary studies. These papers studied several research questions, and described multiple hybrid work settings in agile organizations including work arrangements, online tools, hybrid work policies, as well as agile practices used. 

Even though our mapping study shows a rising trend of research in the intersectional area of hybrid work and agile software development, especially after the Covid-19 pandemic, the number of studies is still low, despite the fact that hybrid work is of high interest for various companies practicing agile software development. Thus, there is a clear need for empirical research on hybrid work in agile software development organizations to be able to support them to thrive in the new working environments.

\bibliographystyleS{unsrt}
\bibliographyS{hybridwork}

\bibliographystyle{ACM-Reference-Format}
\bibliography{hybridwork}

\end{document}